# Analysis of a wet scrubber network in the air remediation of industrial workplaces: benefit for the city air quality

Alessandro Avveduto[1,*], Cesare Dari Salisburgo[1], Lorenzo Pace[1], Gabriele Curci[2,3], Alessio Monaco[3], Marina De Giovanni[1], Franco Giammaria[3], Giuseppe Spanto[4], Paolo Tripodi[1,4]

[1] Società Progetti Innovativi – S.P.In Srl – Via Carlo D'Andrea snc, 67100 Bazzano, L'Aquila

[2] CETEMPS, Centre of Excellence for the integration of remote sensing techniques and numerical modelling for the forecast of severe weather, University of L'Aquila, L'Aquila, Italy

[3] Dept. of Physical and Chemical Sciences, University of L'Aquila, L'Aquila, Italy

[4] Innovation in Sciences&Technologies – Is TECH Srl – Via Mar della Cina 304, 00144 Roma

[*] Corresponding author: alessandro.avveduto@spintecnologia.com

**Abstract**

Industrial activities carried out in confined spaces are characterized by a very specific type of air pollution. The extended exposure to this kind of pollution is often highly harmful, resulting in dramatic effects both on health and safety aspects. The indoor industrial abatement systems, adopted to purify the air, are typically applied to the emission points. The processed air is subsequently emitted outside. In this study we present the experimental results of three-stage wet scrubber systems installed in the industrial workplace of a (i) fiberglass processing plant, where the highest exposure levels to volatile compounds are nowadays today monitored, and of a (ii) waste-to-energy plant, characterized by a very high particulate matter level. The adopted technology, to be used as complementing strategy, does not require special disposal procedures and the processed air is re-emitted in the same work environment for the benefit of the work operators.

The operation of the scrubbers network during the working activities, in both the examined environments, has a) reduced the level of volatile compounds by about 40% and b) maintained the particulate matter concentration PM > 1 to a very low level.

Industrial plants are disseminated in mega-cities and represent dangerous diffuse pollution emission sources. Remediation of highly polluted industrial plants inside our towns, especially for the conventionally untreated emissions, is of extreme benefit for the cities's air quality.

## Introduction

In recent decades, the economic growth and the industrial development have been accompanied by an expansion of the urban area population and by the emergence of megacities. This rapid growth has resulted in a continuously increasing demand for shelter, resources, energy and utilities. At the same time urban emission of air pollutants has grown rapidly, leading to a worsening of the air quality both in the cities and in their surrounding areas (Chan and Yao, 2008).

As is common knowledge, the urban atmospheric pollution is the result of a mixture of chemical compounds and particulate matter, emitted from a variety of anthropic activities such as transportations, residential heating, manufacturing and industrial activities. These latter, in particular, often concentrated in densely populated areas in the largest urban districts (Mcdonald, K., 2012), may represent a significant source of emissions (Cortés et al., 2014).

According to the European Environment Agency (EEA, 2013a), 90% of the urban population in Europe is exposed to pollutant levels above the threshold considered harmful by the World Health Organization (WHO), with a high cost in terms of health, safety and environmental damage (EEA, 2013b). Moreover, in a recent study, long-term exposure to fine particulate matter air pollution has been associated with natural-cause mortality, even for concentrations well below the present European annual mean limit value (Beelen et al., 2013). The need to achieve sustainability in urban environments has never been more acute.

Cut of the emissions, application of new technologies, improved efficiency: these are just some of the recipes usually promoted to mitigate industrial pollutant emissions. But few attention is still given, to this day, to the industrial workplaces environment, where workers spend, on average, about one third of day exposed to a very high level of untreated harmful pollutants, resulting often in serious effects both on health and safety (Kralikova et al., 2014).

The indoor industrial abatement systems adopted to purify the air in the workplaces are typically applied to specific emission points. More frequently, the only caution to mitigate workplace air pollution problems is reduced to the recirculation of air coming from outside, further affecting, consequently, the urban air quality close to the industrial plant where people live densely.

In this study, we report on a first experiments aimed at testing a possible remediation strategy by using a network of Air Pollution Abatement (APA) wet scrubber systems. The proposed method consists in setting up a network of economically and energetically sustainable pollution "absorbers" inside the industrial workplaces, in order to lower pollutant levels at breathable height. In particular the results of the applications inside the workplace of a municipal urban solid waste incinerator and a fiberglass processing plant both located in industrial cities of the northern Italy will be presented.

## Short characterization of the Air Pollution Abatement (APA) wet scrubber

The air cleaner system employed in this study is a prototype of a small three-stage wet scrubber characterized by low consumption, easily operating and adaptable to be installed in network in confined and urban areas. Currently, the scrubber processes about 2300 $m^3$/h of ambient air, requiring a power of 550 W and producing a noise of 55 dBA/1m. Basically, the wet scrubber removes the particles from the air flow incorporating them in liquid droplets, typically water, generated by means of sprayer systems, or in wet plates. In the case of gaseous pollutants the abatement takes place by means of processes of dissolution or absorption of the gas in the liquid. The collected waste material is water, containing all the atmospheric substances captured in the process, to be treated according to the environmental regulations (industrial wastewater management system or municipal water treatment plants). In Fig. 1 is shown the operation scheme of the system. The first stage is composed by three weak Venturi ejectors, efficient in the capture of particulate matter (PM) larger than 1 µm in diameter. The main parameters that influence the PM capturing process are the particle and droplet sizes and the relative velocities between them. In particular, the capture efficiency increases with reducing droplet sizes, with increasing their density and with a higher relative velocity in the throat between the air flow and the drops of liquid (in other terms an increase of the pressure drop between the incoming and outgoing flow). In the second stage a spray ring generates around the container a high concentration of small liquid drops in order to maximize the exchange between the incoming pollutants flow and the liquid phase. It is indicated for the abatement of gas pollutants and coarse particulate. The capture process of gas takes place by means of processes of dissolution in the liquid drops. For this reason the solubility of the gas in the liquid is required (if necessary it is possible to add some additives to the liquid in order to increase the gas solubility). The aerodynamic configuration in this stage must be such that ensure a high degree of mixing between the two phases for a residence time long enough to ensure a satisfactory dissolution of the gas in the liquid. Finally, the third stage is a variable deposition stack where the air flow is forced through a series of wet plates with properly arranged holes. The air flow undergoes a series of alternating compressions and rarefactions phases in order to force the deposition process of pollutants onto the wet plates surfaces.

**Description of the experimental workplace sites**

The experimentation was carried out in two industrial workplaces: the departments of a small artisan business devoted to manufacturing of fiberglass bodywork parts and the furnace room of a municipal solid waste incinerator, Fig. 2.
A fiberglass workplace is characterized by a high level of carcinogenic volatile compounds, such as styrene and epoxy resins (IARC, 1987), and harmful airborne glass-fibers (ACGIH, 2001) (Hoet and Lauwerys, 1992). The remediation action occurred in the two operating departments, *Large* and *Small* room, dedicated to the processing and finishing of fiberglass products, respectively. Two APA wet scrubber systems were positioned near the workstations as shown in Fig. 2(a). The *Large* department, rectangular in shape, includes the finishing department (*Small* room). The *Small* department is approximately 200 m$^3$ in volume, which is about ⅓ of the *Large* one. It is possible to access to the *Small* space only through the *Large* one.

Incinerators reduce urban waste volumes and recover energy by combustion, producing important quantities of solid wastes: bottom ashes (combustion residues) and fly ashes. Studies evidence the presence of toxic heavy metals (Jung et al., 2004) and persistent carcinogenic organic compounds (Durmusoglu et al., 2006), both in bottom and in fly ash. The air remediation occurred in a line of 10 MW of a municipal solid waste incinerator. In particular, it occurred in the rear zone of the furnace, characterized by high exposures of particulate matter coming primarily from the extraction operation of the ashes from the combustion chamber. Two APA wet scrubber systems were installed in front of the ashes extractor, as shown in Fig.2(b). The working activities are performed in an area of about 180 m$^3$ around the rear of the furnace.

No specific air treatment system was installed in both the examined workplaces; only periodic general replacement of the air (window ventilation) occurred.
Measurements of particulate matter concentrations were performed in the workplaces by an AeroTrak Particle Counter Model 9306 produced and calibrated by TSI. The instrument, an *Optical Particle Counter* light-scattering based, detects six dimensional classes in the range 0.30 - 25μm. The level of alcohol gas concentration has been monitored in the large room of fiberglass plant, close to the resin impregnation workstation, by an mq-3 Semiconductor sensor for alcohol gas monitoring in the range 0.04-4 mg/l. The instruments was located in the middle of the experimentation areas, as shown in Fig. 2, at about 150 cm from the ground.

**Results**

The same remediation strategy was applied in both experimental tests maintaining operative h 24 the wet scrubber systems for about two weeks. In order to detect an effect of the wet scrubbers on particulate matter concentrations the air monitoring was carried out for about four weeks during the same working hours characterized by the normal operation of the plant according to logistic and safety constraints. Only in the last two weeks the wet scrubber systems were active.

In Fig. 3 are reported for the six granulometric optical channels the particle concentration monitored in the furnace area of the urban solid waste incinerator. In time series are shown the raw data referring to the periods wet scrubbers not active (APA off, red circle) and active (APA on, green circle) corresponding at about 15 working hours, respectively, characterized by the full operation of the power line (furnace on). The concentration measurements are averaged over one minute collecting an air volume equal to 1.50 liters and provides the particle concentration in different size ranges: [0.30-0.49 μm] [0.50-0.99 μm], [1.00-2.99 μm], [3.00-4.99 μm], [5.00-9.99 μm], [>10.00 μm]. A short period following switching off of the furnace, occurred after the remediation action, is also reported. In total 35 h of air detection were completed over a 25-days period. Considerable variations in the set data, almost up to one order of magnitude, are shown in Fig. 3. The wet scrubbers operation affects visibly on the particle concentration, except in the smaller sizes where small but significant variations are observed. After the remediation period an increasing in particle concentration is monitored, despite the no operation of the furnace, due to working activity performed to restore the line. The results of the experimentation are reported in Fig. 4(a) and refer only to "furnace on" state. The percentage comparison between the average particle concentration measured at the different granulometric channels during scrubbers not active (red bars) and active (green bars) periods is shown. The mean values referred to the no operation period of the air cleaner systems are normalized to 100 at every particle size. Averaging on working hours of different days are taken into account all the possible typologies of activities carried out in the worplace area, making possible the correct quantification of the scrubber action. The effect of the air cleaner systems on workplace air quality occurs for all the investigated granulometric sizes with an average decreasing on particulate matter concentration ranging between 22% [0.30-0.49 μm] and 65%

[>10.00 μm]. In particular, the coarse [> 2.50 μm] and fine [0.10-2.50 μm] particulate is reduced, on average, by about 64% and 32% respectively.

In Fig. 4(b) the particle mean number distribution monitored before (red bars) and during (green bars) the remediation action in the furnace zone is compared with a typical outdoor urban case referred to a big industrial city in the northern Italy (blue bars) (Lollobrigida et al., 2008). The city in question presents similar characteristic to the town where the incinerator is implanted. The reported urban distribution has been averaged over about 300 days of particle detections carried out in a site characterized by traffic flows of the order of 12000 vehicles/day. As expected, the coarse particulate matter concentration is lower in the urban area (blue bars) respect to the mean number concentration monitored in the furnace area (red bars) of the incinerator characterized by high levels of bottom ashes. On the contrary for the fine particulate matter.

During the operation of the APA scrubber systems, the concentration of both fine and coarse particulate matter in the workplace of the furnace zone (green bars), on average, was 50% lower compared to the typical big city environment. It is shown that workers, on average, spend 10.5% of the available time for work activities (Schwanen and Dijst, 2002), often exposed to very high level of untreated harmful pollutants. In this context, the improving of the air quality in industrial workplaces by simple and sustainable remediation strategies (like the one here reported), designed to purify the air at breathable level higher than that of the typical urban environment, would result in savings in terms of health and safety costs and limit the environmental impact on the city (mainly in the area surrounding the industrial plants). According to a recent study (Bertazzi et al., 2007), in fact, an increase of only 10 μg/m$^3$ in the inhaled annual average PM10 produces a variation of +1.94% in deaths and +1.45% in hospitalizations.

Similar results on particulate matter have been found in the experimentation carried out in the operating departments of fiberglass plant but are not shown here. In this study we report the effect of wet scrubber systems on the $C_nH_m$ compounds inside the processing fiberglass department, *Large room*, characterized by high levels of volatile chemical compounds. In Fig. 5 we compare $C_nH_m$ measurements (3 minutes average) during two working days characterized by the same activities. The blue and red lines are referred to the period wet scrubber systems not active and active, respectively. The use of a.u. is motivated by confidential issues. In this test the APA scrubber systems were maintained operative in the workplace only during the working hours (8:00-19:00). The emissions consequent to working production start about at 9:00 a.m., one hour after the plant opening, and persist oscillating until the stop activities at 19:00. Subsequently, it is observed an increase in $C_nH_m$ concentration as a result of the emissions from the heat treatment stage of the fiberglass finished products: the manufacts are maintained at a temperature of about 65° for one or two hours and fluxed with air by an open recirculation system. The highest concentration reached in this stage is approximately three times the average concentration achieved during the working hours and occurred about one hour after the closure of the working area. Finally, after the heat treatment stage, the $C_nH_m$ compound concentration decreases exponentially because of the deposition process and reaches in about three hours an equilibrium value. The $C_nH_m$ concentration is relatively constant (200 a.u.) during the night hours (0:00 - 8:00 a.m.). This behavior is always observed during the experimentation. When the APA system is activated at 8:00 a.m., the $C_nH_m$ concentration (red line), on average, drops in about one hour (before the starting of working activities) to a level of about 30% lower than the no operation period of the scrubber (blue line). Moreover, during the working activities, the scrubber action keeps the concentration of $C_nH_m$ compounds constantly lower than the no operation period. When APA is turned off, at 19:00, the $C_nH_m$ concentration monitored during the remediation action (red line) follows the same behavior observed before the air treatment (blue line), but with values roughly halved. Finally, the $C_nH_m$ concentration averaged on the working times 8:00 - 19:00 during the scrubber action is about 40% lower than that measured during the no operation of the APA system.

**Conclusions**

The identification and promotion of alternative technological solutions able to minimize the pollutants exposures in specific industrial workplaces is required (Türschmann et al., 2011), everywhere in mega cities where industrial plants are disseminated in the urban tissue representing harmful diffuse emission sources
In this study we have reported the effect of sustainable pollution "absorbers", based on a three stage wet scrubber technology, inside two industrial workplaces, implanted in urban area, characterized by the presence of toxic particulate matter and persistent organic volatile compounds
Our results suggest that sustainable and simple industrial remediation strategies could contribute significantly to improve workplace air quality and limit the industrial environmental impact on urban area for the benefit of both workers and the urban population at large.

# Figures

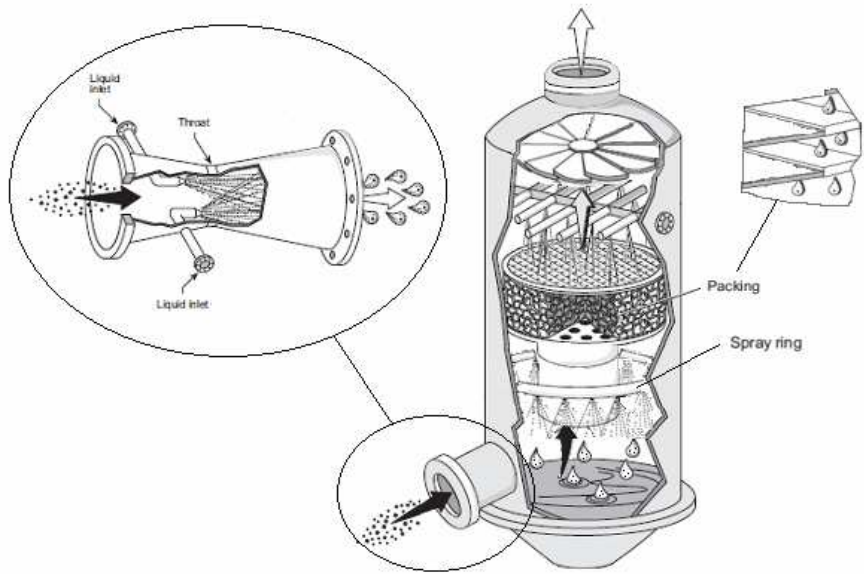

Fig. 1. Operation scheme of the Air Pollution Abatement (APA) three-stage wet scrubber system. The first stage is composed by weak Venturi ejectors. The second stage is performing a spray tower with an appropriate water spray ring. Finally, the third stage is a variable deposition stack, packed with a series of proper wet plates.

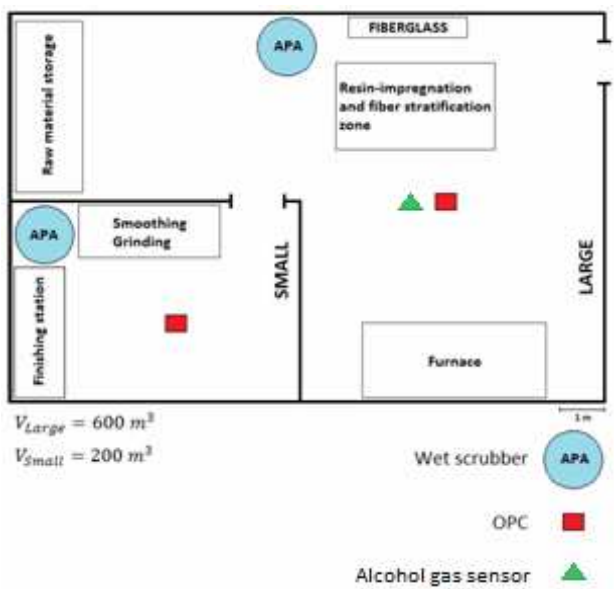

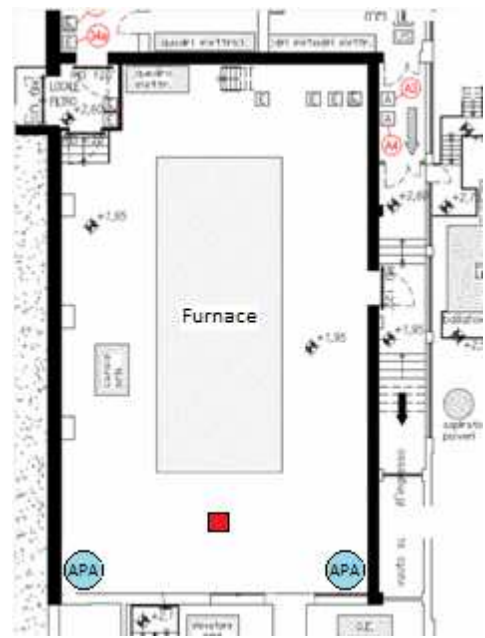

(b)

(a)
Fig. 2. Experimental workplace sites. Manufacturing departments of fiberglass processing plant (a) and furnace room of municipal solid waste incinerator (b). The location of the APA wet scrubbers and the sampling points of particulate matter and alcohol gas are also shown.

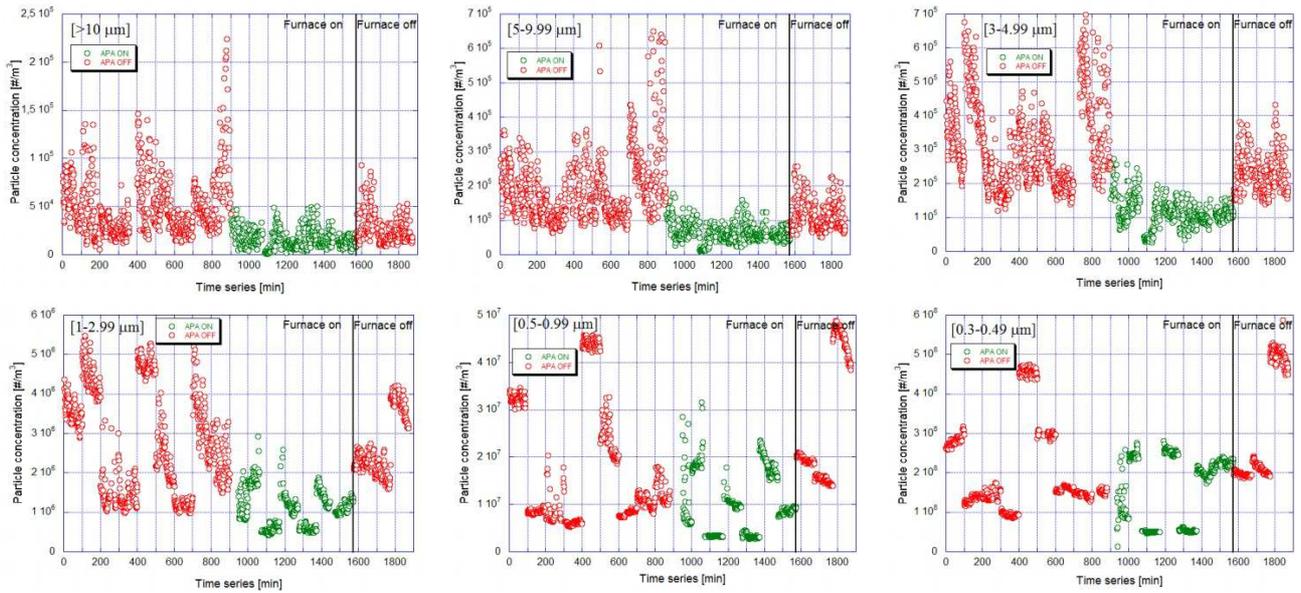

Fig. 3. Time series of the particle concentration measurements performed by *OPC* at six granulometric channels in the furnace room of the solid waste incinerator. Red and green circles represent the concentration values monitored during the periods not active and active scrubbers, respectively. The operating status of the furnace is also indicated.

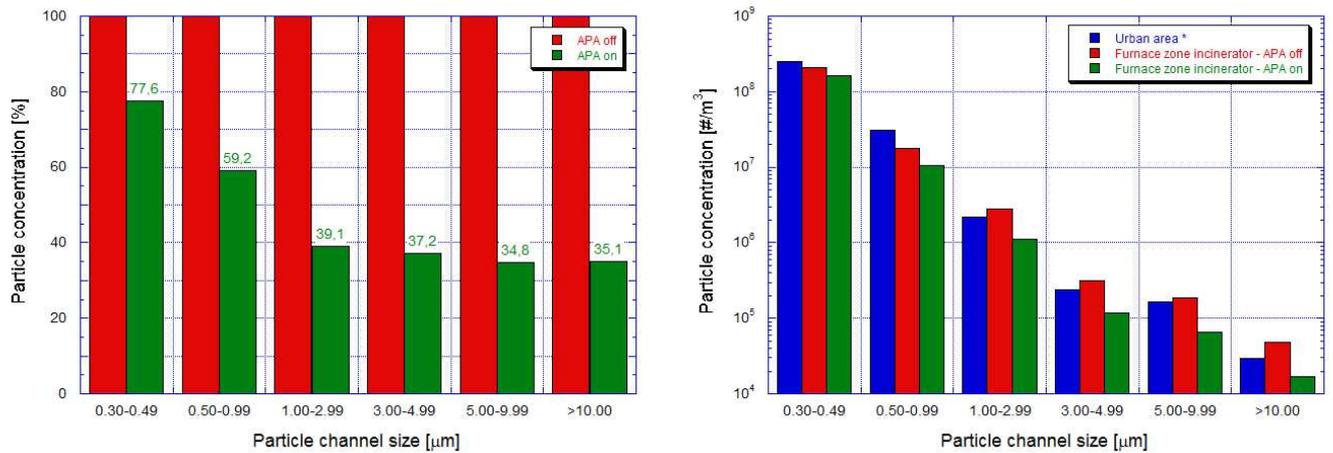

(a)                                                                                                                              (b)

Fig. 4. (a) Comparison between the average particle concentration measured at different granulometric channels during the periods not active (red bars) and active (green bars) scrubbers in the furnace room of the solid waste incinerator. The APA off mean values at every particle size are normalized to 100. (b) Particle number distribution monitored in the furnace room during APA off (red bars) and on (green bars) periods, respectively, compared with a typical urban particle distribution (blue bars) *(Lollobrigida et al., 2008).

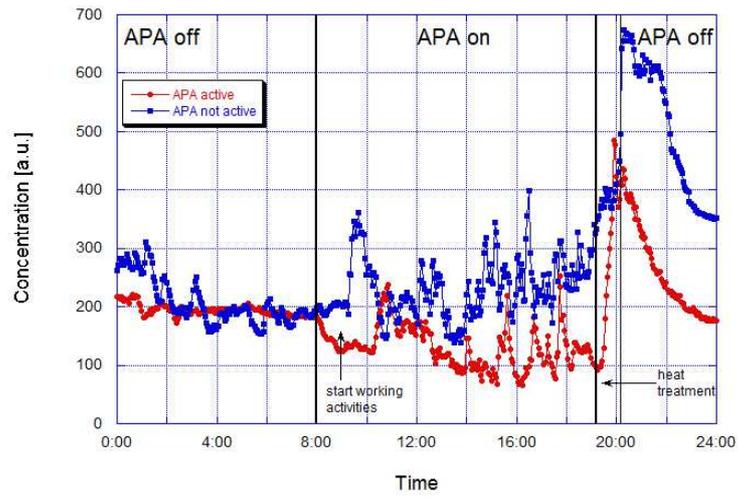

Fig. 5. $C_nH_m$ compounds concentration measured by *MQ-3* Alcohol gas sensor in the *Large* workplace of the fiberglass plant in two working days during the period APA not active (blue points) and active (red points, scrubber operating only from 8:00 to 19:00), respectively.